\documentclass[apj]{emulateapj}

\newcommand\hMpc{$h^{-1}{\ }{\rm Mpc}$}
\newcommand\hMsun{$h^{-1}{\ }{\rm M_{\odot}}$}

\newcommand\Rvir{$R_{\rm vir}$}
\newcommand\Mvir{$M_{\rm vir}$}

\newcommand\zform{$z_{\rm form}$}
\newcommand\sizeg{0.7}
\newcommand\degrees[1]{\ensuremath{#1^\circ}}

\begin{document}

\title{Radial Alignment in Simulated Clusters}
\author{Maria J. Pereira, Greg L. Bryan and Stuart P. D. Gill}
\affil{Columbia University, Department of Astronomy, New York, NY, 10025}
\email{pereira@astro.columbia.edu}
\begin{abstract}
Observational evidence for the radial alignment of satellites with their dark matter host has been accumulating steadily in the past few years. The effect is seen over a wide range of scales, from massive clusters of galaxies down to galaxy-sized systems, yet the underlying physical mechanism has still not been established. To this end, we have carried out a detailed analysis of the shapes and orientations of dark matter substructures in high-resolution N-body cosmological simulations. We find a strong tendency for radial alignment of the substructure with its host halo: the distribution of halo major axes is very anisotropic, with the majority pointing towards the center of mass of the host. The alignment peaks once the sub-halo has passed the virial radius of the host for the first time, but is not subsequently diluted, even after the halos have gone through as many as four pericentric passages. This evidence points to the existence of a very rapid dynamical mechanism acting on these systems and we argue that tidal torquing throughout their orbits is the most likely candidate. 
\end{abstract}


\section{Introduction}

Anisotropy in galaxy orientations has been a matter of debate for several decades, and many conflicting reports have been published. Past studies have found evidence for three different types of alignment: alignment between clusters \citep{bin82,pli02}, between the brightest cluster galaxy (BCG) and the satellite distribution \citep[e.g.,][]{yan06} and between the orientation of satellites and their host \citep{haw75,per05,agu06}. This last type of alignment, which we will refer to as \emph{radial} alignment and is the focus of this paper, has been the hardest to confirm \citep{tre92,tor07}, since it requires high quality data on small scales. In recent years, this field has seen a resurgence, largely due to the arrival of the Sloan Digital Sky Survey (SDSS) \citep{aba05}. The SDSS provides accurate measurements of isophotal shapes for millions of galaxies, and this has finally allowed large statistical studies of galaxy alignments to be performed. \citet{per05} targeted galaxies in massive X-ray selected clusters and found a significant tendency for their radial alignment. This result has since been confirmed by \citet{fal07a} for a larger sample of groups optically selected from the SDSS. On smaller scales, \citet{agu06} found a tendency for satellite galaxies in the SDSS to be radially aligned with their host galaxy, whereas on large scales, \citet{man06} found a very significant correlation between the orientations of galaxies and the surrounding density field traced by galaxy overdensities. 

Initially motivated by the prospect of using galaxy orientations to probe their formation histories, these studies are now also driven by the need to calibrate weak lensing and cosmic shear measurements. A key assumption for lensing techniques is that the population of galaxies being lensed is randomly oriented. Some intrinsic alignments between galaxies can be dealt with easily: \emph{e.g.} downweighting close pairs readily removes contamination by alignments induced in interacting systems. However, as \citet{hir03} pointed out, if galaxy orientations are affected by their surrounding density field  (\emph{e.g.} a galaxy cluster), then they will also be correlated with the orientations of the background population of galaxies that is being lensed by that field. This correlation between widely separated redshift bins cannot trivially be removed.

Given the growing body of evidence suggesting that galaxy orientations are anisotropic, and the pressing need for an accurate quantification of intrinsic alignments for weak lensing, it seems crucial that we try and find the physical cause behind these anisotropies. There are two commonly proposed explanations. The first, initially developed by \citet{pee69} in his tidal torque theory (TTT), explains the anisotropy as a left-over primordial effect. TTT ascribes the orientation and rotation of galaxies to torquing during their formation. It therefore follows that the signal should be stronger on the outskirts of the cluster, and that it wanes with time, such that older, more relaxed clusters  should exhibit less tendency for alignment. The other alternative, proposed by \citet{per05}, is a dynamical mechanism, i.e. an interaction with the tidal field of the host cluster that gets progressively stronger during infall and is not erased by subsequent orbital motions. Observational studies have so far been unable to distinguish between the two, due to difficulties in constraining galaxy orbits and in measuring galaxy shapes accurately out to large redshifts. 

A different approach is needed, and a few numerical studies have recently been published on this subject. Studies of simulated halo shapes and orientations have been performed around voids \citep{bru07}, along filaments and sheets \citep{ara07,alt06,hah07} and in a Milky-Way type halo \citep{kuh07}. Anisotropies are reported in all three environments. The advantages of working with simulated clusters are obvious - 3D spatial information means we do not suffer dilution from projection effects. Also, with enough temporal and spatial resolution, we can follow the galaxies as they fall into the cluster along filaments, and beyond, as they orbit inside the cluster. By tracking the effect's evolution with time, we will be able to more precisely determine its source.

We start (\S 2.1) by introducing the simulations used for this analysis and describing the properties of the eight host halos. Our methods for finding the substructure halos (\S 2.2) and determining their shapes (\S  2.3) follow, along with a study of the reliability of our shape measurements. With this information, we then show in section 3.1 that cosmological dark-matter simulations do indeed produce radial alignment in clusters at z $\approx 0$, and we study the correlation of this effect with various parameters, such as host halo mass and distance from the cluster center (\S 3.2). Having established the importance of the alignment effect in our simulations,  we use the high temporal resolution to study its evolution with time in section 3.3, and its dependence on orbital phase (\S 3.4). We argue in section 4.1 that tidal torquing by the host halo tidal field is responsible for the alignment of substructure, and compare our results with previous observations in \S 4.2. We end (\S 4.3) by briefly speculating on the possible consequences of such a mechanism for the morphological and orbital evolution of galaxies in clusters.

\section{Simulations and Analysis}

\subsection{The Data}
The $N$-body simulations used in this work are presented in detail in \citet{gil04}, and we describe them here only briefly. Using the open source adaptive mesh refinement code \texttt{MLAPM}
\citep{kne01}, a set of four initial conditions at redshift
$z=45$ in a standard $\Lambda$CDM cosmology ($\Omega_0 = 0.3,\Omega_\lambda =
0.7, \Omega_b h^2 = 0.04, h = 0.7, \sigma_8 = 0.9$) were created.  From an initial distribution of $512^{3}$ particles
 in a box  64\hMpc\ wide and with a mass resolution of
$m_p = 1.6 \times 10^{8}$\hMsun, the closest eight particles were iteratively collapsed, reducing the particle number to 128$^3$ particles. These low resolution
initial conditions were then evolved until $z=0$, at which point eight clusters were selected in the mass range  1--3$\times 10^{14}$\hMsun. All particles within two times the virial radius were then tracked back to their initial positions at $z=45$, where they were regenerated to their original mass resolution and positions. These high resolution pockets are surrounded by a ``buffer" zone with eight times the original mass resolution, which itself is nested in particles that are 64 times more massive than the particles at the center of the cluster. These initial conditions were then re-simulated to $z=0$, recording 63 outputs from $z=1.5$ to $z=0$ so that $\Delta t \approx 0.17$ Gyrs. A summary of the eight host halos is presented in Table~\ref{HaloDetails}, and on quick inspection it should be immediately apparent that the eight hosts have widely varying masses and assembly histories. We calculate these quantities as follows: the virial radius is defined as the distance at which the average halo density drops below $\rho_{halo}(r_{vir}) = \Delta_{vir}\rho_b$, where $\Delta_{vir} = 340$ and $\rho_b$ is the cosmological background density. The virial mass is defined to be the mass inside this radius. We calculate each host's age as the time elapsed since their formation, which is defined, following \citet{lac93}, at the redshift where the halo first contains half of its present-day mass.

\begin{table}[h]
\begin{center}
\begin{tabular}{cccccc}\hline
Halo & \Rvir &  \Mvir  & \zform & age &
$N_{\rm sat}(<r_{\rm vir})$ \\
 
\hline \hline
  \# 1 & 1.34 & 	   2.82	 & 	1.18   & 	8.37	 & 		166 \\
\# 2 & 0.97 & 	   1.05	 & 	0.87   & 	7.17	 & 		49  \\
\# 3 & 1.06 & 	   1.38	 & 	0.84   &  	7.01	 & 		98  \\
\# 4 & 1.06 & 	   1.38	 & 	0.75   & 	6.57	 & 		71  \\
\# 5 & 1.34 &      2.80  & 	0.59   & 	5.65	 & 		168 \\
\# 6 & 1.06 & 	   1.39	 & 	0.50   & 	5.06	 & 		97  \\
\# 7 & 1.00 & 	   1.16	 & 	0.43   & 	4.52	 & 		54  \\
\# 8 & 1.37 & 	   3.00	 & 	0.30   & 	3.42	 & 		152 \\
\hline
\end{tabular}
\end{center}
\caption {\label{HaloDetails}Summary of the eight host dark matter halos at $z=0$. Distance is measured
         in \hMpc, mass in 10$^{14}$\hMsun, and age in Gyrs. Only satellites with more than 200 particles are tallied in the last column.}

\end{table}

\subsection{Identifying Substructure}

Simulation outputs merely tell us what the particle spatial and kinetic distribution is at each redshift. They give us no information about particle assignment or halo identity - which particle belongs to which halo? 
There is no unique answer to this question, mainly because there are many different ways in which a halo can be defined. 

A number of sophisticated algorithms have been developed to locate halos within simulations \citep{dav85,fre88,ber91,sut92,wei97,kly97}. They face many challenges: the dynamic environment of cosmological simulations blurs halo boundaries, and halos are continually undergoing mergers or being stripped within a host potential, making it impossible to clearly define a halo edge. Furthermore, most of these do a poor job at finding substructure in very dense background regions, and although nearly all algorithms now use kinetic information to remove gravitationally unbound particles, they are generally not too concerned with background contamination, which can be safely ignored for most applications.

Unfortunately, these issues are especially problematic for our analysis: once the substructure halos cross the virial radius of the host cluster, contrast is lost, and the particle background from the host becomes very significant. If we were to mistakenly assign background particles from the cluster to our substructure halos, this could mimic the radial alignment effect we are looking for, since cluster particles themselves are radially distributed. We solve this problem by finding the substructure halos early on, at the formation redshift ($z_{form}$) for each host. At these early times, the hosts are still starting to assemble and halos are not as clustered, and therefore much easier to identify. Once the halos have been found, their individual particle distributions can then be tracked forward in time through any environment, even into the densest cores of clusters, without suffering from background contamination. 

A more detailed description of our halo finding and tracking methods can be found in \citet{gil04}, so we provide here only a brief summary. We find and truncate all the halos in our simulation volume at $z_{form}$ using the AMIGA Halo Finder (AHF), the successor of the MLAPM Halo Finder (MHF) \citep{gil04}. AHF uses the adaptive grids of AMIGA to locate halos within the simulation. AMIGA's adaptive refinement meshes follow the density distribution \it by construction\rm. The grid structure naturally ``surrounds'' the halos, as the halos are simply manifestations of overdensities. As AMIGA's grids are adaptive it constructs a series of embedded grids, the higher refinement grids being subsets of grids on lower refinement levels. AHF takes this hierarchy of nested isolated grids and constructs a ``grid tree''. Within that tree, each branch represents a halo, thus identifying halos, sub-halos, sub-sub-halos and so on.

Once we have found all the halos and sub-halos in our simulations at this redshift, we can start tracking their particle distributions through time. The main disadvantage of this method is that any subsequent accretion (after $z_{form}$) onto the halos will, by design, be ignored. This seems a reasonable compromise - halos within halos travel through their environment too quickly to accrete significant amounts of particles and we assume that any particles acquired before the halo enters the host settle into the potential well isotropically, so that we are still obtaining a fair sample of the shape of the halos by only including the particles that were present at $z_{form}$.  

At each time step we look at the distribution of particles for each halo and, after recalculating their center of mass, we check if each particle is still bound to the halo. This is an iterative process: starting at the center of the halo and moving outwards, we calculate each particle's kinetic and potential energy in the halo's reference frame and remove all particles that have velocities, $v > b v_{esc}$, where $b = 1.5$ is the bound factor, and the only free parameter in our algorithm. We repeat the process until no further particles are removed or a minimum number ($N_p = 200$, \emph{c.f.} \S 2.3) of particles has been reached. Particles that are determined to be unbound are subsequently ignored. This is a completely effective way of removing the cluster background, as particles that do not belong to the substructure halo will be quickly left behind. It also allows us to track debris being stripped off the subhalos as they orbit inside the cluster. 

When all unbound particles have been removed, we fit an NFW distribution to the radial profile of the remaining particles. We define the halo's radius as the distance at which the average halo density drops below $\rho_{halo}(r_{vir}) = \Delta_{vir}(z)\rho_b(z)$, where $\Delta_{vir}(z)$ is the virial overdensity at that redshift, and discard any particles that lie outside this limit. However, this radius is almost never reached in the case of substructure, in which case the radius of the halo is defined as the distance to the furthest bound particle. Once we have determined which particles belong to which halo at each timestep, we are ready to measure their shapes.

\subsection{Shape Measurements}

How can we condense a three dimensional particle distribution into a few simple parameters describing its shape? With no prior knowledge of how the particles are distributed this is a difficult task. However, halos produced in dark matter cosmological simulations seem to follow a universal density profile \citep{nfw96}, and are generally well fit by triaxial ellipsoids \citep{fre88, all06}. The simplest way to do this is to calculate the inertia tensor of the distribution, $I_{jk} = \sum_i m_i r_{i,j} r_{i,k}$, which is then diagonalized to find the principal axes of the halo. However, this procedure is not ideal, since it weights particles by $r^2$, and therefore results in a shape measurement that is overly biased by the outlying particle distribution. 

A better measure \citep{ger83} is the \emph{reduced} inertia tensor:
\begin{equation}\label{eq inertia tensor}
\tilde{I}_{jk} = \sum_i m_i \frac{ r_{i,j} r_{i,k} }{r_i^2}.
\end{equation}
which weights particles equally regardless of their distance to the center of the halo, using only the directional information of halo particles to calculate their shapes. The eigenvectors and eigenvalues of this reduced form of the inertia tensor give us the principal axes of the halo and a measure of their relative lengths ($b/a, c/a$), although the latter are substantially overestimated, as we shall see.

The main source of uncertainty in determining the shapes of our halos is the small number of particles we sample their potentials with. We want to characterize halo alignments over as wide a mass range as possible, so we want to know what the minimum number of particles is that will still give us a reliable measure of a halo's shape. The ability to determine the orientation of a halo's major axis also depends strongly on the value of $b/a$: An oblate halo with $b \approx a$, will be almost degenerate in its major/intermediate axis orientations. 

\begin{figure}[h]
\begin{center}
\scalebox{\sizeg}{\includegraphics{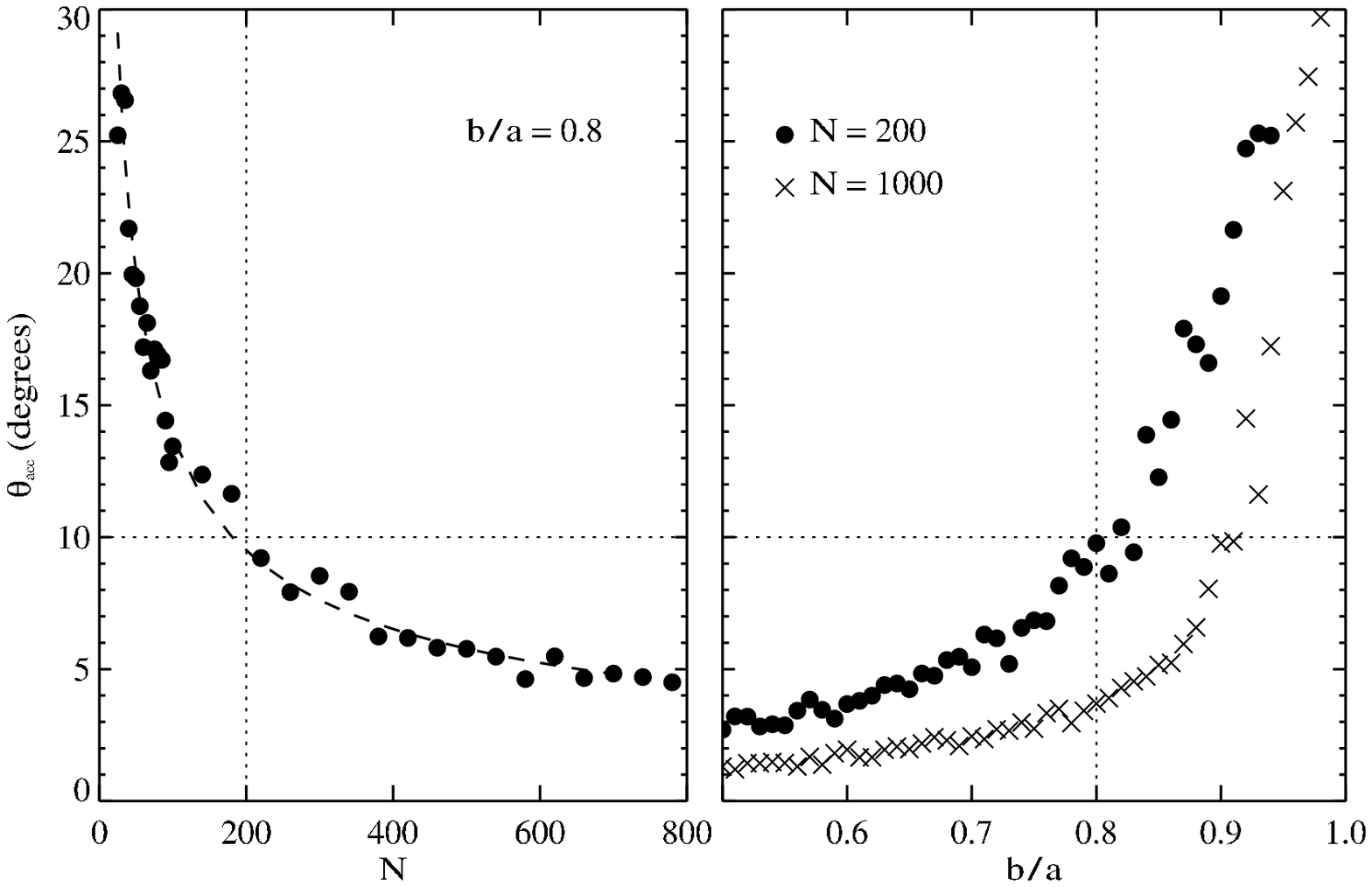}}
\caption{\footnotesize{\label{err1} Results for ``fake" NFW halo analysis. \emph{Left:} Error in major axis orientation vs. number of particles in halo. The vertical dashed line represents the lower limit on N for the halos analysed in this paper, whereas the horizontal lines show the minimum accuracy that will be tolerated. \emph{Right: } The error in major axis orientation vs. the intermediate-to-major axis ratio. Errors are shown for halos with 200 particles (as dots) and with 1000 particles (as crosses).}}
\end{center}
\end{figure}

In order to address these questions, we generated a set of fake triaxial NFW halos with varying numbers of particles ($N_p$) and intermediate-to-major axis ratios ($b/a$) and fed them through our pipeline.  For each value of $N_p$ and $b/a$ we performed 100 random realizations of an NFW halo and calculated the angle between the major axis direction measured and that which was input. The dispersion in these values, $\theta_{acc}$, is then a good estimate of the accuracy of our measurement. The minor-to-major axis ratio ($c/a$) does not appear to affect the determination of the major axis direction, and the results presented in figure \ref{err1} are therefore only for prolate halos with $b = c$.

As expected, our accuracy depends very strongly on the number of particles sampled - the points on the left panel are well fit by a relation of the form:  $\theta_{acc} \propto N^{-0.54}$. We want a compromise between individual halo accuracy and sample size - at values of $N_p<200$, $\theta_{acc}$ increases rapidly, and we  pick this, somewhat arbitrarily, for our lower limit on $N_p$. If $b/a = 0.8$, our measurements of these halos would be accurate by $\theta_{acc} \approx \degrees{10} $. However, $\theta_{acc}$ also depends strongly on $b/a$. When $N_p = 200$, $b/a < 0.8$ is required to maintain the \degrees{10} error, with an increase in $b/a$ leading to a rapid decrease in accuracy. Figure \ref{err1} refers to the input values of $b/a$. In fact, the measured ellipticities are much higher, although the two are tightly correlated: $(b/a)_{input}=(b/a)^{0.45}_{measured}$. We place an upper limit of 0.8 on the intrinsic axis ratios, which translates to a limit on the measured values of $b/a < 0.9$.

With our limits in place for the minimum number of particles and maximum axis ratios, we are ready to start analyzing our results. It is worth noting, however, that both these error sources would bias our shapes randomly: there is no preferred direction that will be selected if the halos are under-sampled or too spherical. This in turn implies that the results on alignment presented in the next section are, if anything, conservative.

\section{Results}
\label{res}
\subsection{Alignment at $z=0$}
\label{z0sec}
\begin{figure}[h]
\begin{center}
\scalebox{\sizeg}{\includegraphics{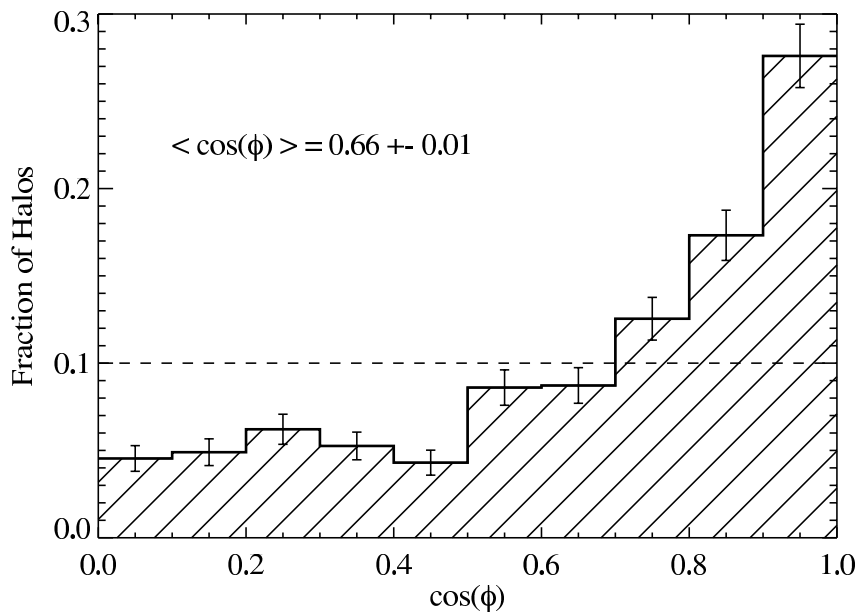}}
\caption{\footnotesize{\label{z0}Angle distribution at $z=0$. An isotropic distribution 
would follow the dashed line, with $\left< \cos{\phi} \right> = 0.5$}}
\end{center}
\end{figure}

The quantity we will focus on is the angle, $\phi$, between the major axis of each halo and the vector connecting the halo to the center of the host. If halos are oriented randomly in space, the cosine of  $ \phi$ will be uniformly distributed between 0 and 1, with a mean value, $\left< \cos{\phi} \right>$, of 0.5. When $\cos \phi \approx 1$ the halo is pointing toward the host center, whereas when $\cos \phi \approx 0$ it is aligned tangentially to it, so that a value of $\left< \cos{\phi} \right> > 0.5$ implies an overall tendency for radial alignment. The standard error on $\left< \cos{\phi} \right>$ is $\sigma_{\left< \cos{\phi} \right>} = \sigma/\sqrt{N}$, where $N$ is the sample size and $\sigma$ is its standard deviation. We show in Figure \ref{z0} a histogram of  $\cos{\phi}$ for all halos within $2 r_{vir}$ of each of the eight hosts. It is immediately apparent that our distribution is inconsistent with isotropy at a very high significance level: $\left< \cos{\phi} \right> = 0.66 \pm 0.01$, with most halos pointing toward the center of the host halo.

\begin{figure}[h]
\begin{center}
\scalebox{\sizeg}{\includegraphics{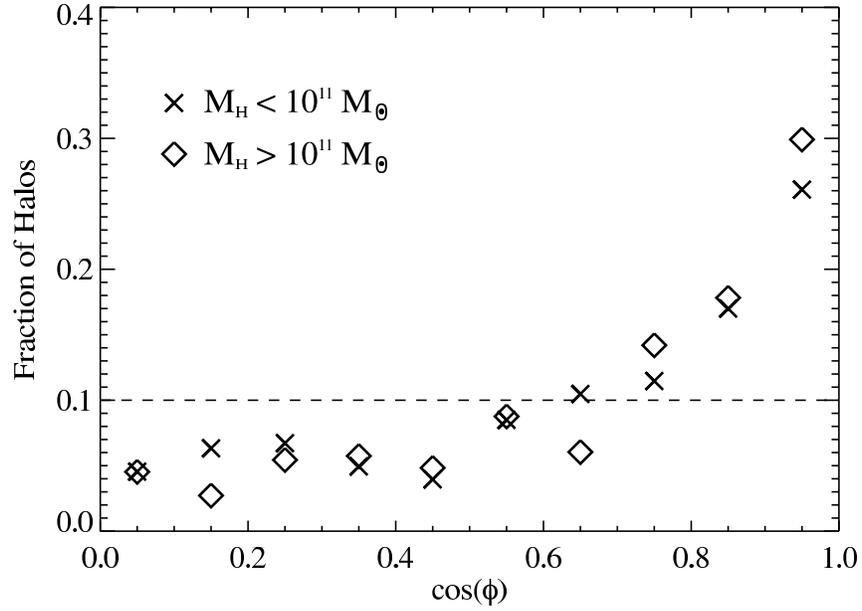}}
\caption{\footnotesize{\label{z0mass}Angle distribution at $z=0$ for high mass and low mass halos. An isotropic distribution 
would follow the dashed line, with $\left< \cos{\phi} \right> = 0.5$}. Mass limits are in units of $M_{\odot}$.}
\end{center}
\end{figure}

While it is clear that the results of  figure \ref{z0} confirm previous observational reports of radial alignment,  a precise quantitative comparison is rather difficult, and we defer this discussion to \S \ref{obs}. Nevertheless, much can be learned from a qualitative study of the effect's behavior and correlation with individual (and host) halo properties. 
Figure \ref{z0mass} shows the same histogram as in Figure \ref{z0} but now for two separate halo populations, segregated by mass. There does not appear to be a significant distinction between the two populations. This tells us not only that the alignment effect is mass independent, but also confirms the experiments in  \S 2.3 that show that resolution effects are unimportant in the lowest mass halos considered in our analysis ($N_p>200$).

We can also study how the effect depends on extrinsic characteristics of the halo, \emph{e.g.} the distance to the center of the host, or the host mass. We searched for correlations with different global host properties such as mass and age, and found none. The alignment mechanism appears to be universal, in that it is present with approximately the same strength in hosts with widely varying mass, formation times and assembly histories. This surprising result is also seen in observational studies: \citet{per05} found no correlation of the alignment strength with the dynamical state of the clusters inferred from their x-ray morphologies.

\subsection{Dependence on Distance to Cluster Center}

\begin{figure}[h]
\begin{center}
\scalebox{\sizeg}{\includegraphics{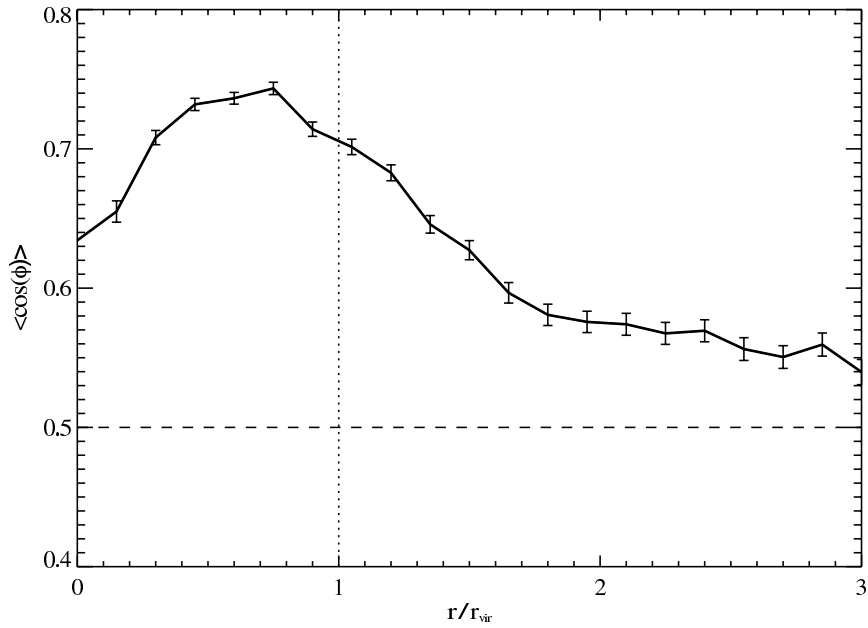}}
\caption{\footnotesize{\label{z0dist}Radial alignment vs. distance to cluster center for all halos with $z<z_{form}$. An isotropic distribution is again represented by the dashed line}}
\end{center}
\end{figure}

Figure \ref{z0dist} shows the dependence of the effect on the distance from the cluster center. All halos  at redshifts $z<z_{form}$ are included in this analysis, in order to enhance the overall signal. The behavior appears very smooth: $\left< \cos{\phi} \right>$ rises gradually as the host is approached, peaks slightly past its virial radius, and then decreases again toward the center. 

It is striking that already at a distance of three virial radii there is a small, consistent, tendency for radial alignment. At this distance, how can the halo already ``feel" the presence of the host? This is easily understood once we consider that clusters form at the intersection of filaments, and hence that most filaments will be radially aligned with respect to their nearest clusters. If there is a primordial alignment of halos with respect to the filaments in which they form, then even at large distances this will appear as a radial alignment in our analysis. This type of primordial alignment at large radii was seen by \citet{ara07} in their study of filamentary structures, where they found similar values for $\left< \cos{\phi} \right>$ (their figure 2e).  

The main focus of this paper, however, is what happens closer to the host. As the halo falls in, the amplitude of the alignment increases dramatically, reaching a peak of $\left< \cos{\phi} \right> = 0.72$ at about one-half of the virial radius of the cluster, before decreasing again gradually inside the core. 
The increase of the alignment with decreasing distance matches the behaviour found by \citet{fal07a} in their study of SDSS groups, and is to be expected if the effect is caused by the tidal field of the host, but the dip at small radii, $r < 0.3r_{vir}$, has not yet been observed. This is most likely due to the severe projection effects that dominate the cores of observed clusters.
 
What causes this behavior?  It appears that the alignment that is set-up in the infall regions is being disrupted in the inner regions of the cluster. What causes the disruption? Is this primarily a spatial effect caused by the environment of the cluster core, or a temporal one, given that galaxies closest to the center have been in the cluster environment for longer? And what produced the alignment in the first place? The best way to answer these questions is to take advantage of the extra dimension provided by simulations, and explore the evolution of this effect with time.
 
\label{distsec}

\subsection{Evolution with Redshift}

\begin{figure}[h]
\begin{center}
\scalebox{\sizeg}{\includegraphics{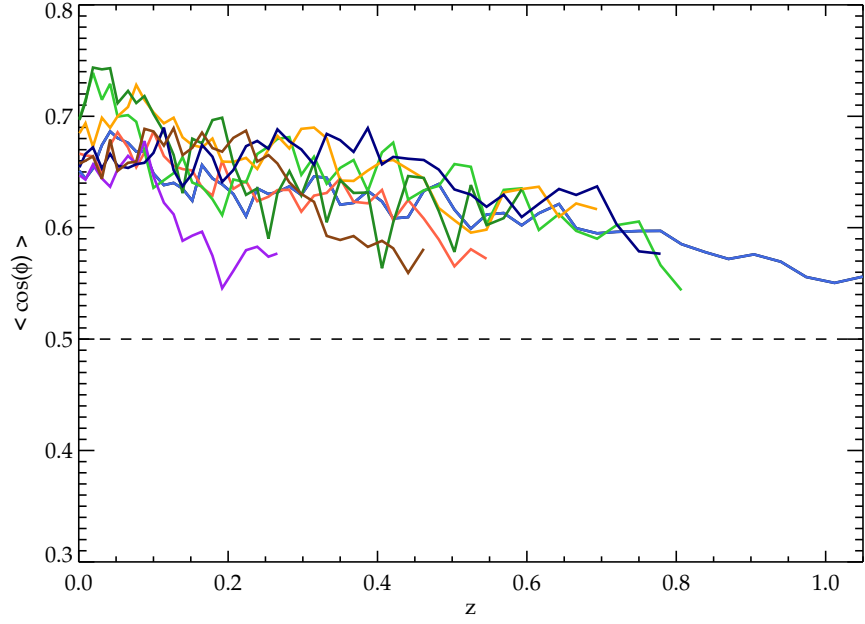}}
\caption{\label{zall8}\footnotesize{Evolution of radial alignment with redshift for each of the eight clusters. Each cluster is plotted for $z < z_{form}$, and the alignment signal at each redfshift is averaged over every halo within $r_{vir}$.}}
\end{center}
\end{figure}

The evolution of the alignment with redshift is plotted in figure \ref{zall8} for each of the eight clusters independently. 
Perhaps the most striking feature of this plot is the self-similarity of the different curves. Every cluster appears to go through exactly the same evolution, regardless of size or formation time, such that at $z=0$ they are practically indistinguishable, as described in the previous section.
Figure \ref{zall8} also reveals that the clusters evolve monotonically, with the strength of the effect increasing steadily since the formation time of each cluster to the present day.
  
Whatever the source of the disruption at the cluster cores, it is seemingly not strong enough to dilute the overall alignment signal. There are two possible explanations: It could be that, even though alignment is disrupted once the halo reaches the core of the host, the constant infall of pristinely aligned halos results in an overall increase of the average alignment per host. Alternatively, the misalignment seen at the cores could be short-lived - a feature of each halo's orbital motion through the potential of the host. 

Distinguishing between these two alternatives requires a different approach: we need to track halos on their way toward the cluster, and then trace their orbits inside the virial radius of the host.

\subsection{Evolution with Orbital Phase}

\begin{figure}[h]
\begin{center}
\scalebox{\sizeg}{\includegraphics{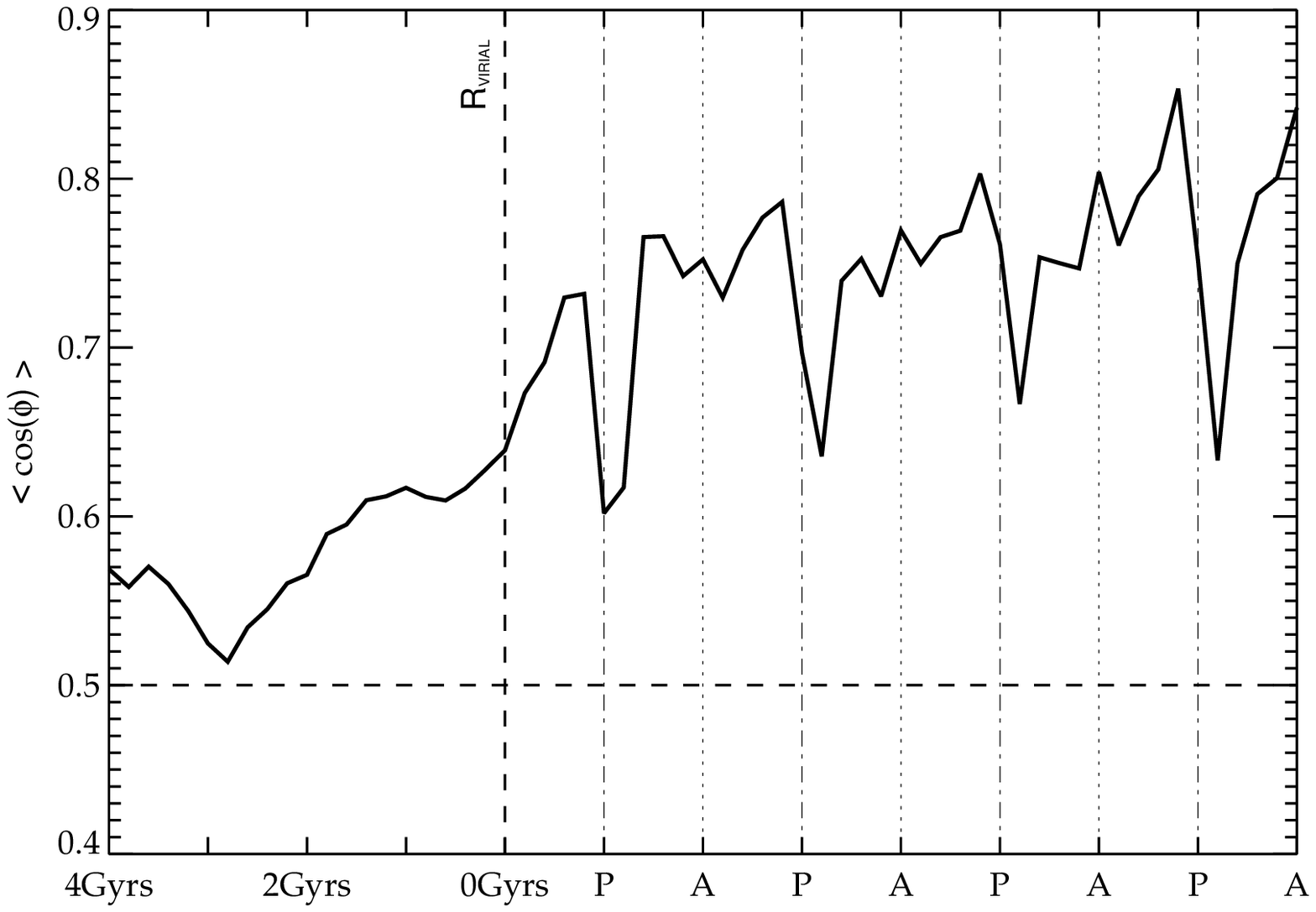}}
\caption{\label{orbit}\footnotesize{ Average radial alignment vs. orbital phase for all sub- 
halos. The horizontal axis represents time - initially in Gyrs before crossing host R$_{vir}$ and 
then in subsequent Pericentric (P) and Apocentric (A) passages. }}
\end{center}
\end{figure}

Figure \ref{orbit} shows the alignment evolution stacked for all halos throughout their orbits. Initially halos are tracked relative to the amount of time (in Gyrs) remaining until they cross the virial radius of the host for the first time. Once they cross this threshold, halo orbital times are normalized at each passage through pericenter and apocenter. 

We again detect a small alignment at large distances from the cluster , which we believe is evidence for a primordial alignment along filaments as discussed in \S \ref{distsec}. As the host is approached the signal increases significantly, peaking just before the first pericentric passage, and then a periodic oscillation ensues, which follows the halo's orbital period closely. On average, the tendency for alignment is much larger within the host than before, although the alignment tendency changes dramatically with orbital phase. It follows that the dip observed near the cluster cores in figure \ref{z0dist} is in fact a result of the misalignment observed at pericenter, and, most importantly, that it is not disruptive, since the alignment tendency is restored well before the next apocenter is reached. In fact, the alignment is quite constant throughout the rest of the orbit  and seems to increase slightly at each passage. This evidence points to a stable dynamical effect that is set-up as the halo orbits around the cluster. 

\begin{figure}[h]
\begin{center}
\scalebox{\sizeg}{\includegraphics{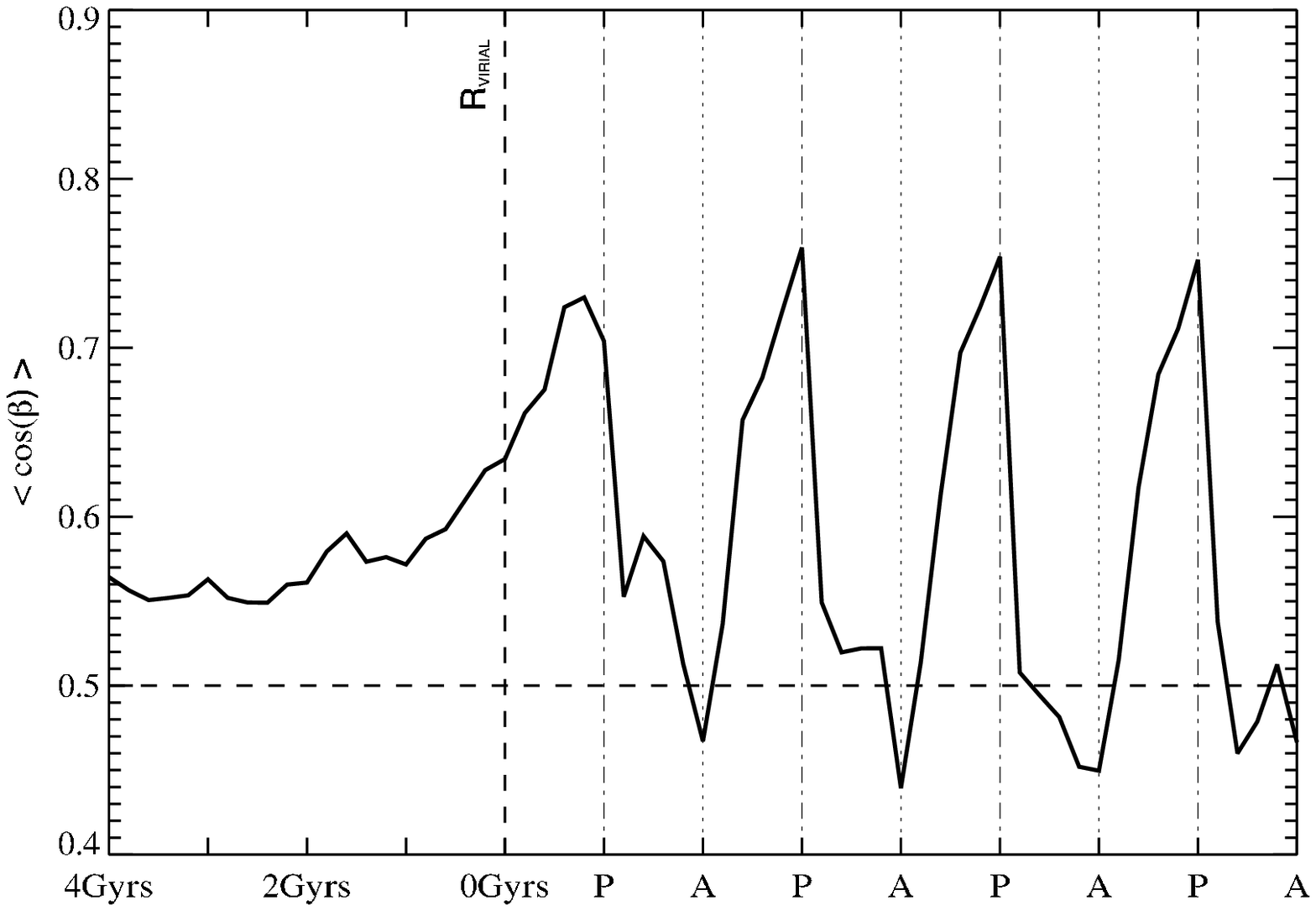}}
\caption{\label{orbal}\footnotesize{ Average orbital alignment vs. orbital phase for all sub- 
halos. The horizontal axis represents time - initially in Gyrs before crossing host R$_{vir}$ and 
then in subsequent Pericentric (P) and Apocentric (A) passages.}}
\end{center}
\end{figure}

Further insights can be obtained by exploring the orientations of the halos with respect to their orbits. We define a new angle, $\beta$, as the angle between each halo's major axis and the halo's velocity, and plot  the mean value of its cosine for all halos vs. orbital phase in Figure \ref{orbal}. The similarities in behavior between radial and orbital alignment at large distances are simply a consequence of the radial nature of the orbits themselves - halos form and travel along filaments toward the intersecting nodes where clusters reside. In fact, even inside the hosts, orbits are quite eccentric, with an average apocentric to pericentric distance ratio of $4:1$. 

Could it be, then, that the radial alignment we observe within the virial radius is just a tendency for halos to be aligned along their orbits coupled with the fact that orbits are, on average, quite radial? Figure \ref{orbal} tells us that this is not the case: once inside the cluster, we find that the orbital alignment is also correlated with orbital phase, but whereas the radial alignment is almost instantly recovered after pericenter, the orbital alignment increases again much more slowly, and only after reaching the next apocenter. This asymmetry around pericenter seems, at first, surprising, but, as will be shown in the following section, follows as a natural consequence of tidal torquing by the cluster potential throughout the halo's orbit.

\section{Discussion}
\subsection{Tidal Torquing as a Mechanism for Alignment}
\label{tor}
\begin{figure}[h]
\begin{center}
\scalebox{\sizeg}{\includegraphics{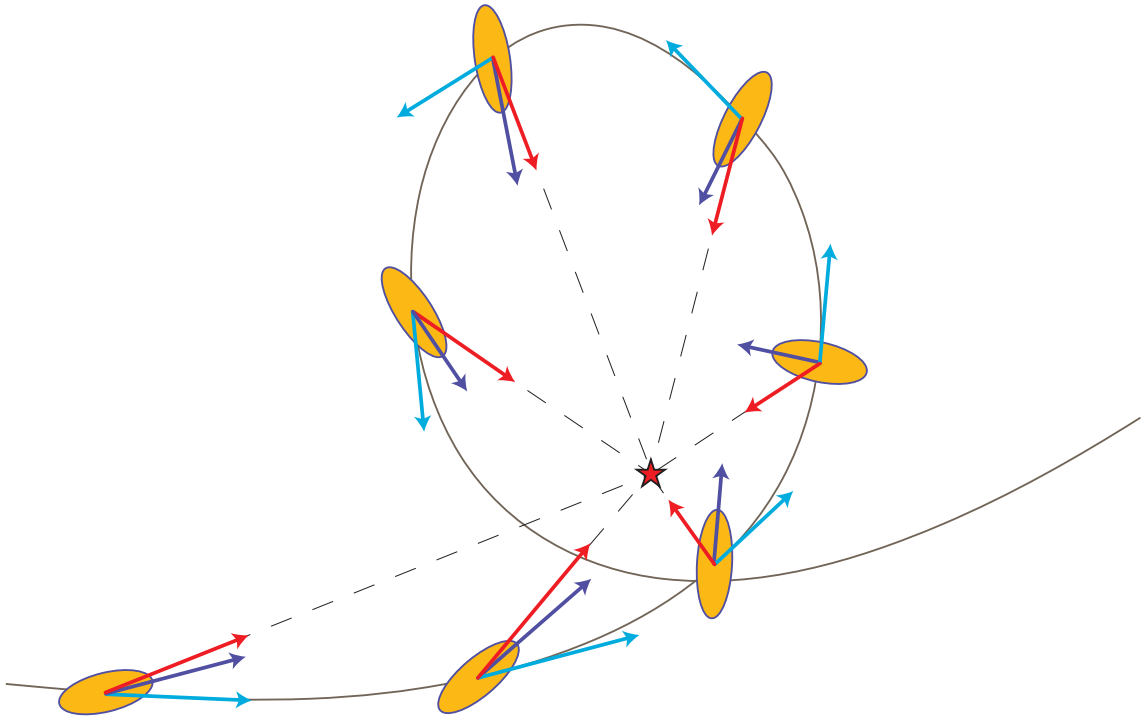}}
\vspace{5 mm}
\caption{\footnotesize{\label{sketch}A sketch of a representative halo orbit around its host. The center of mass of the host is represented by a star. The radial direction is drawn as a red vector, cyan indicates the orbital direction, and the purple vector is the direction of the major axis of the halo. Purple and red vectors are generally very close, apart from a short mismatch at pericenter caused by the high orbital velocities. Red and cyan vectors are close before pericenter, but almost orthogonal to each other in the second part of the orbit.}}
\end{center}
\end{figure}

Once radial and orbital alignment information is combined, a clearer picture emerges of what is going on inside these clusters. As the halo approaches pericenter along an eccentric orbit, it is continually torqued along the direction of the potential gradient, i.e. halos tend to point toward the host center, and, because their orbits are fairly eccentric, also along their orbital direction. At pericenter, the halo is moving too fast for the torquing to be completely effective, which causes the dip in radial alignment. It is nevertheless enough to torque the halo away from its orbital direction and back toward the cluster center, in a figure rotation that is co-planar with its orbital rotation and in the same direction. The radial alignment is quickly reinstated, but orbital alignment is lost as the halo progresses towards apocenter. Steady torquing throughout the orbit keeps halos oriented toward the cluster center and away from the direction of their orbits until after the apocentric passage, where orbital alignment increases steadily towards pericenter, and a new cycle begins. Figure \ref{sketch} illustrates this behaviour with a sketch of a halo's rotation as it orbits around the cluster.

	If halo orbits were circular, halos would quickly become tidally locked and maintain radial alignment throughout their orbits. In reality, their orbits are quite eccentric, and their orbital speed varies significantly. Halos do not react to the tidal torquing quickly enough through the pericentric passage, and the narrow dips observed are the result. In fact, idealized numerical experiments involving a single halo in a circular orbit around a static host invariably lead to tidal locking of the halo, although the time required for locking varies significantly with the original orientation of the halo (C.\ M.\ Simpson \& K.\ V.\ Johnston, private communication). Interestingly, for halos that start out already pointing toward the host center, the time required is rather short, of the order of an orbital period or less.

\begin{figure}[h]
\begin{center}
\scalebox{\sizeg}{\includegraphics{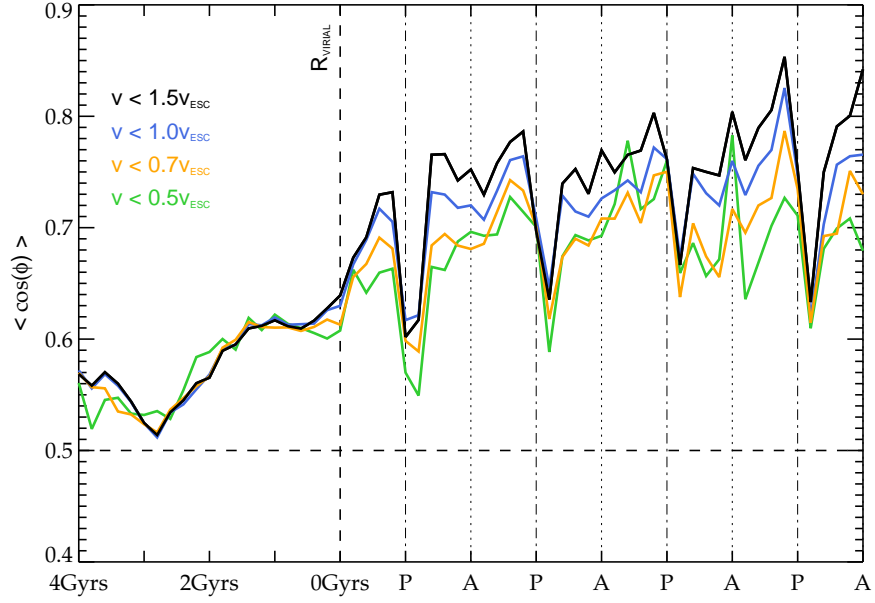}}
\caption{\label{orbitb}\footnotesize{ Same as figure \ref{orbit} but for varying values of $b$, the particle binding factor ($b=v/v_{esc}$). $b$ ranges from the original value of 1.5 (least bound, in black) to 0.5  (most bound, in green). }}
\end{center}
\end{figure}

Further support for this tidal torquing hypothesis is shown in figure \ref{orbitb}. Although we believe our shape measurements to be robust to random outliers, it is possible that strongly distorted outer shells, caused, e.g., by tidal stripping, could significantly bias the result. As a test, we apply four different particle cuts to each of our halos by varying the boundedness criteria on the particle velocities: instead of throwing out all particles for which $v>b v_{esc}$, where $b = 1.5$,  we exclude alternately particles that have velocities greater than 1, 0.75 and 0.5 times the escape velocity. For the most conservative criteria, which only retains particles that have velocities $v < 0.5 v_{esc}$, more than $70\%$ of the particles are discarded, and we are only probing the very bound cores of the halos. Figure \ref{orbitb} makes clear that stripping cannot possibly be the sole cause of the alignment effect, since even the most conservative cut shows significant alignment. 

Nevertheless, a trend is observed, in that the most bound particles show slightly less tendency for alignment overall. This could be the result of tidal stripping in the outer layers, but more likely it is a simple statistical effect:  Because we only consider halos with $N_p>200$ in this analysis, as we progressively exclude more particles from the halo with decreasing $b$, some halos fall below this limit and are consequently ignored. Hence a decrease in $b$ implies a smaller number of halos in each sampled bin, which reduces the signal-to-noise, and brings $\left< \cos{\phi} \right>$ closer to $0.5$.  
Despite this trend, the conclusion remains that halo shapes are not significantly warped by tidal stripping, and that tidal torquing of the entire halo is a better explanation for the effect.


A number of early numerical and analytical studies support the importance of tidal torques within clusters . \citet{mil82} performed a set of numerical experiments on a rotating bar in an external force field and observed tidal braking of the rotation, with a rate that was inversely related to the square of the cluster crossing time. More recently, numerical $N$-body experiments by \citet{cio94} showed that the time required for the alignment of a prolate galaxy with the tidal field of a cluster is much shorter than the Hubble time, and on the order of a few times the galaxy's intrinsic dynamical time. Using a different approach, \citet{usa97} studied tidal effects on gaseous ellipsoids orbiting in a central potential analytically, predicting that galaxies in eccentric orbits should have their long-axis trapped toward the direction of the radius vector of the cluster.

While this paper was being written, two studies were published on halo alignments that describe similar results. \citet{kuh07} studied the alignment of substructure around a Milky Way type halo using the Via Lactea simulation, and observed a radial alignment tendency that is preserved throughout the halos' orbits. \citet{fal07b} looked at several different types of alignment in a set of dark matter hosts at $z=0$, finding similar levels of radial alignment that increase with decreasing distance to the host.

\subsection{A Comparison with Observations}
\label{obs}
The results of \S \ref{res} certainly seem to substantiate the observational evidence for radial alignment of cluster galaxies. A quantitative comparison, however, is not easily made.
In order to properly ``observe" these simulations, semi-analytic models of galaxy formation are required to extrapolate from the dark matter halos to the luminous components embedded within. These then need to be projected, interlopers and survey limits accounted for, and the resulting image fed through traditional source extraction and isophotal analysis pipelines. This is a laborious procedure and it cannot yet provide accurate results, since galaxy formation models are still largely unconstrained in a crucial parameter: the alignment between luminous and dark matter.
 
Observationally, studies of the alignment and relative ellipticity of the two components are currently only possible for gravitational lens galaxies, a very rare class of objects.  \citet{kee98} analysed a sample of 17 lenses, mostly isolated early-types, and found that the luminous component of the lens generally aligns with its inner halo to $\le$ \degrees{10}. In order to probe the shapes of the halos to larger radii, stacked galaxy-galaxy weak lensing studies are needed. These are just now becoming feasible, and preliminary results appear somewhat contradictory \citep{hoe04,man06b}.

Most theoretical studies have concentrated on the formation of disk galaxies and their angular momentum, where some misalignment between baryonic and dark matter spin is commonly seen (e.g. \citet{bos03}). On the other hand, \citet{bai05} find that the orientations of simulated halos and their embedded disks are largely uncorrelated at large radii, and almost perfectly aligned at small ($r< 0.1r_{vir}$).

The current uncertainty in this parameter makes it impossible to accurately predict the orientation of the galaxies that would populate our halos. However, the results presented in this paper suggest a gravitational origin for the alignment mechanism, and it is therefore reasonable to expect that the two components should react similarly to it. Furthermore, the tidal torquing within clusters is so effective that the halos appear to ``forget" their original orientations before a single orbit is completed, which renders the original alignment between light and dark matter relatively unimportant.

\begin{figure}[h]
\begin{center}
\scalebox{\sizeg}{\includegraphics{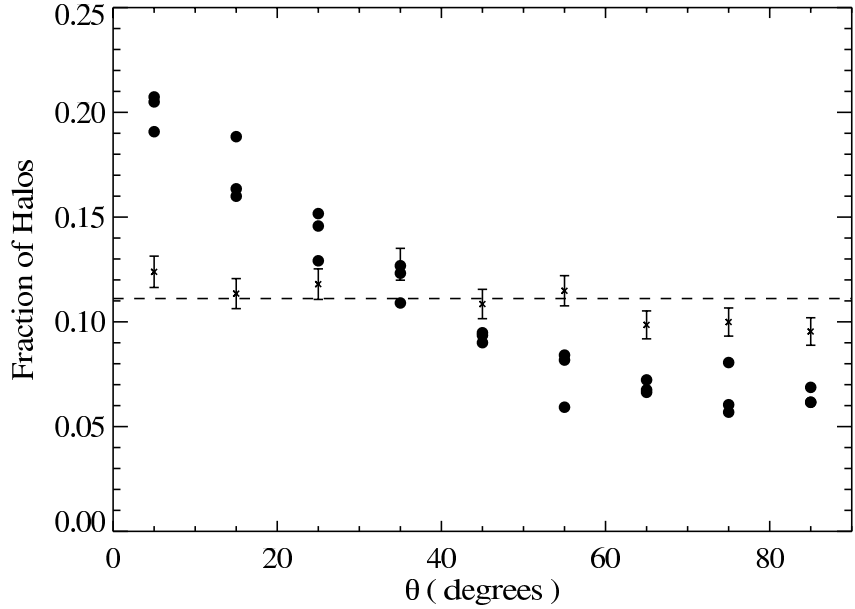}}
\caption{\footnotesize{\label{2d}2D radial angle distribution at $z=0$ for all halos within $2 r_{vir}$ of the cluster centers, projected along the three independent spatial dimensions (filled circles). Also plotted for comparison is the observed angle distribution from \citet{per05}, with Poissonian error bars.  An isotropic distribution would follow the horizontal dashed line, with $\left< \theta \right> = \degrees{45}$}}.
\end{center}
\end{figure}

We therefore compare the dark matter alignment directly with the galaxy observations of \citet{per05}. 
We project each halo's dark matter particles along the three spatial dimensions in our simulation and compute the 2D inertia tensor of their projected distribution. The angle between the halo's 2D major axis and its projected separation from the cluster center can then be measured. We include in this sample all galaxies within 2 virial radii of the cluster center - interlopers are not accounted for, since the SDSS galaxies we are comparing our results to are all spectroscopically confirmed cluster members.  Figure \ref{2d} shows the results of this 2D analysis and compares them with the SDSS observations.  
We plot all three independent projections and note that the dispersion in their values should give us a fair estimate of the error introduced by the projection procedure itself. 

The dark matter alignment is much stronger than that observed: $\left< \theta \right> _{halo} = \degrees{34}.5 \pm \degrees{0}.9$ whereas $\left< \theta \right> _{gal} = \degrees{42}.79 \pm \degrees{0}.55$.
This must be in some part a reflection of how much harder it is to measure accurate galaxy position angles on an survey image than for a well-resolved halo in a cosmological simulation, where dynamical information allows for a much cleaner background removal. Nonetheless, the dilution caused by this measurement noise cannot wholly account for the significant difference in radial alignment between the two components. Given the nature of the alignment mechanism established in \S \ref{tor}, it is perhaps not too surprising that dark matter halos are more strongly aligned. One would naively expect the dark matter halos to be more easily torqued, given that they are much more extended (providing a longer lever) and have generally lower spins ( and therefore less gyroscopic resistance) than their luminous counterparts.

\subsection{Possible Consequences of Tidal Torquing in Clusters}
Halo alignments have traditionally been studied either as a probe of their formation history, or as a contaminant to weak lensing studies. Now that we have established that the leading mechanism behind halo alignments within clusters is a dynamical effect present throughout their lifetime, it is interesting to speculate on what possible evolutionary consequences this mechanism might have for the halos affected.

Figure \ref{sketch} shows us that at each point in the orbit, the torque acts to rotate the halo away from its orbital direction, which necessarily results in a deceleration of the halo's orbital motion, inducing orbital decay. The halos analysed in this study do indeed show a tendency for orbital circularization: \citet{gil04b} showed that halos with more pericentric passages have smaller orbital eccentricites. They argued that dynamical friction was not a likely cause, and tentatively ascribed the effect to the growth of the host halo instead. While the velocity dispersion of the satellites is seen to depend on the host halo mass, it seems possible that at least part of the orbital decay observed is a natural result of the constant torquing throughout the halo's orbit. This is an interesting prospect, since an extra source of orbital decay could potentially help solve the outstanding problem of cD formation in massive clusters, as well as alleviate some unresolved discrepancies between observed satellite populations and the generally low efficacy of dynamical friction predicted by numerical studies (\emph{e.g.} \citet{has03,taf03}). We are currently investigating the importance of this induced orbital decay, and this will be the subject of a future paper.

Another possible consequence of the strong torquing of dark matter halos within hosts is the possibility of disk warping. Because of their high angular momentum, disks will naturally resist tidal torquing more effectively than the surrounding dark matter halo, which will introduce a misalignment between the halo and the disk. Even though recent studies of (isolated) disk-halo alignments show that their orientations are largely uncorrelated at large radii \citep{bai05}, the same is not true for the inner halos ($r<0.1r_{vir}$), where the rotational axis of the disk is seen to lie very close to the minor axis of the inner halo. We have shown that tidal torquing affects all particles in the halos, even the most bound, so it is not unreasonable to expect that the inner shells should also feel these torques. The question then remains whether the disk within will align itself accordingly, or whether the misalignment could be a possible cause of warping of the disk, but this will also require further study.

\section{Conclusions}

There is growing observational evidence that a satellite's major axis is preferentially aligned with the radial vector linking the satellite to its host.  This tendency for satellites to point at their hosts has been seen on both cluster and group scales \citep{per05, agu06}.  Motivated by this result, we have used a suite of cosmological N-body simulations to investigate the alignment between satellite and host dark matter halos. 

We take particular care to separate satellite and cluster particles using a combined halo finder plus tracker.  In this method, an adaptive halo finder \citep{gil04} is used to initially identify a set of satellite sub-halos that we subsequently track as they enter and orbit the cluster, removing particles as they become unbound.  The advantage of this approach is that we can be sure to use only genuine sub-halo particles and exclude ``background" cluster particles that might bias our shape measurements.  We then use the reduced inertia tensor to measure the shapes and orientations of all sub-halos which end up inside the virial radii of a set of eight simulated clusters.  We highlight here the main results obtained from this analysis:
 
\begin{itemize}

\item The satellites in the simulations show a strong tendency to point toward the cluster center.  The mean cosine of the angle between the major axis of each halo and the cluster center is $\left< \cos{\phi} \right> = 0.66 \pm 0.01$, where an isotropic distribution would have $\left< \cos{\phi} \right> = 0.5$.  This tendency for alignment is found for all clusters at all redshifts analyzed, and does not appear to depend on the mass of the cluster or the satellite.

\item The amplitude of the alignment is a strong function of radius, with a small but significant effect extending out to many virial radii from the cluster.  This signal, which has been seen in previous work \citep{ara07}, is most likely left over from the primordial imprint of the surrounding large-scale structure and can be ascribed to tidal torques exerted at early times, when the cluster-size perturbations were just turning around \citep[e.g.,][]{pee69}.

\item Closer to the cluster center, within 1-2 virial radii, the amplitude of the alignment increases dramatically to a peak of $\left< \cos{\phi} \right> = 0.72$ at about one-half of the virial radius, and then falls slowly closer to the cluster center.  

\item When examined as a function of orbital phase for a given satellite, we find that the alignment increases rapidly as the satellite falls into the cluster for the first time and remains high after that, except for a short period during pericenter passage, when it dips precipitously.  It is this short-lived dip which gives rise to the decrease in $\left< \cos{\phi} \right>$ close to the cluster center.

\end{itemize}

Based on these results, we conclude that the strong alignment seen at small radius --- within two virial radii --- is due to {\em tidal torquing} by the cluster halo. The idea is very simple -- the galaxy is only in a stable equilibrium if it is pointing at the cluster center; otherwise there is a net torque which acts to rotate the galaxy towards this equilibrium point.  We demonstrate that the alignment is seen both in the outer and inner parts of the satellite, indicating that it is not due to some process (such as tidal stripping) which impacts only the outer, poorly bound, part of the sub-halo.  We also briefly review previous literature which has investigated the impact of tidal torques on collisionless systems using analytic approximations or idealized simulations, and find that the expected amplitude and timescale is sufficient to produce the alignments we see.

Although we study only dark-matter halos, we expect this effect to extend to the luminous part of galaxies, as observations seem to indicate.  This will have an observational impact on weak lensing studies and may also modify the distribution of stars in a satellite, as well as the satellite's orbital properties.  We will investigate these possibilities in future work.

\acknowledgements
The simulations presented in this paper were carried out on the Beowulf cluster at the Centre for Astrophysics \& Supercomputing, Swinburne University. We would like to thank Jeff Kuhn, Kathryn Johnston and Christine Simpson for helpful discussions.  Greg Bryan acknowledges support from NSF grants AST-05-07161, AST-05-47823, and AST-06-06959, as well as the National Center for Supercomputing Applications.


\clearpage

\begin{thebibliography}{}
\bibitem[Agustsson \& Brainerd(2006)]{agu06} Agustsson, I., 
\& Brainerd, T.~G.\ 2006, \apjl, 644, L25
\bibitem[Allgood et al.(2006)]{all06} Allgood, B., Flores, 
R.~A., Primack, J.~R., Kravtsov, A.~V., Wechsler, R.~H., Faltenbacher, A., 
\& Bullock, J.~S.\ 2006, \mnras, 367, 1781 
\bibitem[Altay et al.(2006)]{alt06} Altay, G., Colberg, 
J.~M., \& Croft, R.~A.~C.\ 2006, \mnras, 370, 1422 
\bibitem[Arag{\'o}n-Calvo et al.(2007)]{ara07} 
Arag{\'o}n-Calvo, M.~A., van de Weygaert, R., Jones, B.~J.~T., \& van der 
Hulst, J.~M.\ 2007, \apjl, 655, L5  
\bibitem[Abazajian et al.(2005)]{aba05} Abazajian, K., et 
al.\ 2005, \aj, 129, 1755 
\bibitem[Bailin et al.(2005)]{bai05} Bailin, J., et al.\ 
2005, \apjl, 627, L17
\bibitem[Bertschinger \& Gelb(1991)]{ber91} Bertschinger, E., 
\& Gelb, J.~M.\ 1991, Computers in Physics, 5, 164 
\bibitem[Binggeli(1982)]{bin82} Binggeli, B.\ 1982, \aap, 
107, 338 
\bibitem[Brunino et al.(2007)]{bru07} Brunino, R., Trujillo, 
I., Pearce, F.~R., \& Thomas, P.~A.\ 2007, \mnras, 375, 184 
\bibitem[Ciotti \& Dutta(1994)]{cio94} Ciotti, L., \& Dutta, 
S.~N.\ 1994, \mnras, 270, 390 
\bibitem[Davis et al.(1985)]{dav85} Davis, M., Efstathiou, 
G., Frenk, C.~S., \& White, S.~D.~M.\ 1985, \apj, 292, 371 
\bibitem[Djorgovski(1983)]{djo83} Djorgovski, S.\ 1983, 
\apjl, 274, L7 
\bibitem[Faltenbacher et al.(2007)]{fal07a} Faltenbacher, A., 
Li, C., Mao, S., van den Bosch, F.~C., Yang, X., Jing, Y.~P., Pasquali, A., 
\& Mo, H.~J.\ 2007, ArXiv e-prints, 704, arXiv:0704.0674 
\bibitem[Faltenbacher et al.(2007)]{fal07b} Faltenbacher, A., 
Jing, Y.~P., Li, C., Mao, S., Mo, H.~J., Pasquali, A., \& van den Bosch, 
F.~C.\ 2007, ArXiv e-prints, 706, arXiv:0706.0262 
\bibitem[Frenk et al.(1988)]{fre88} Frenk, C.~S., White, 
S.~D.~M., Davis, M., \& Efstathiou, G.\ 1988, \apj, 327, 507 
\bibitem[Gerhard(1983)]{ger83} Gerhard, O.~E.\ 1983, \mnras, 
202, 1159 
\bibitem[Gill et al.(2004)]{gil04} Gill, S.~P.~D., Knebe, A., 
\& Gibson, B.~K.\ 2004, \mnras, 351, 399 
\bibitem[Gill et al.(2004)]{gil04b} Gill, S.~P.~D., Knebe, A., 
Gibson, B.~K., \& Dopita, M.~A.\ 2004, \mnras, 351, 410 
\bibitem[Hahn et al.(2007)]{hah07} Hahn, O., Carollo, C.~M., 
Porciani, C., \& Dekel, A.\ 2007, ArXiv e-prints, 704, arXiv:0704.2595 
\bibitem[Hahn et al.(2007)]{hah07} Hahn, O., Carollo, C.~M., 
Porciani, C., \& Dekel, A.\ 2007, ArXiv e-prints, 704, arXiv:0704.2595 
\bibitem[Hashimoto et al.(2003)]{has03} Hashimoto, Y., 
Funato, Y., \& Makino, J.\ 2003, \apj, 582, 196 
\bibitem[Hawley \& Peebles(1975)]{haw75} Hawley, D.~L., \& 
Peebles, P.~J.~E.\ 1975, \aj, 80, 477 
\bibitem[Hirata \& Seljak(2003)]{hir03} Hirata, C., \& 
Seljak, U.\ 2003, \mnras, 343, 459 
\bibitem[Hoekstra et al.(2004)]{hoe04} Hoekstra, H., Yee, 
H.~K.~C., \& Gladders, M.~D.\ 2004, \apj, 606, 67 
\bibitem[Keeton et al.(1998)]{kee98} Keeton, C.~R., Kochanek, 
C.~S., \& Falco, E.~E.\ 1998, \apj, 509, 561 
\bibitem[Klypin \& Holtzman(1997)]{kly97} Klypin, A., \& 
Holtzman, J.\ 1997, ArXiv Astrophysics e-prints, arXiv:astro-ph/9712217 
\bibitem[Knebe et al.(2001)]{kne01} Knebe, A., Green, A., \& 
Binney, J.\ 2001, \mnras, 325, 845 
\bibitem[Kuhlen et al.(2007)]{kuh07} Kuhlen, M., Diemand, J., 
\& Madau, P.\ 2007, ArXiv e-prints, 705, arXiv:0705.2037 
\bibitem[Lacey \& Cole(1993)]{lac93} Lacey, C., \& Cole, S.\ 
1993, \mnras, 262, 627 
\bibitem[Mandelbaum et al.(2006a)]{man06} Mandelbaum, R., 
Hirata, C.~M., Ishak, M., Seljak, U., \& Brinkmann, J.\ 2006, \mnras, 367, 
611 
\bibitem[Mandelbaum et al.(2006b)]{man06b} Mandelbaum, R., 
Hirata, C.~M., Broderick, T., Seljak, U., \& Brinkmann, J.\ 2006, \mnras, 
370, 1008 
\bibitem[Miller \& Smith(1982)]{mil82} Miller, R.~H., \& 
Smith, B.~F.\ 1982, \apj, 253, 58 
\bibitem[Navarro et al.(1996)]{nfw96} Navarro, J.~F., Frenk, 
C.~S., \& White, S.~D.~M.\ 1996, \apj, 462, 563 
\bibitem[Peebles(1969)]{pee69} Peebles, P.~J.~E.\ 1969, \apj, 
155, 393 
\bibitem[Pereira \& Kuhn(2005)]{per05} Pereira, M.~J., \& 
Kuhn, J.~R.\ 2005, \apjl, 627, L21 
\bibitem[Plionis \& Basilakos(2002)]{pli02} Plionis, M., \& 
Basilakos, S.\ 2002, \mnras, 329, L47 
\bibitem[Suto et al.(1992)]{sut92} Suto, Y., Cen, R., \& 
Ostriker, J.~P.\ 1992, \apj, 395, 1 
\bibitem[Taffoni et al.(2003)]{taf03} Taffoni, G., Mayer, L., 
Colpi, M., \& Governato, F.\ 2003, \mnras, 341, 434 
\bibitem[Trevese et al.(1992)]{tre92} Trevese, D., Cirimele, 
G., \& Flin, P.\ 1992, \aj, 104, 935 
\bibitem[Torlina et al.(2007)]{tor07} Torlina, L., De 
Propris, R., \& West, M.~J.\ 2007, \apjl, 660, L97 
\bibitem[Usami \& Fujimoto(1997)]{usa97} Usami, M., \& 
Fujimoto, M.\ 1997, \apj, 487, 489 
\bibitem[van den Bosch et al.(2002)]{bos03} van den Bosch, 
F.~C., Abel, T., Croft, R.~A.~C., Hernquist, L., \& White, S.~D.~M.\ 2002, 
\apj, 576, 21 
\bibitem[Weinberg et al.(1997)]{wei97} Weinberg, D.~H., 
Hernquist, L., \& Katz, N.\ 1997, \apj, 477, 8 
\bibitem[Wesson(1984)]{wes84} Wesson, P.~S.\ 1984, \aap, 138, 
253 
\bibitem[Yang et al.(2006)]{yan06} Yang, X., van den Bosch, 
F.~C., Mo, H.~J., Mao, S., Kang, X., Weinmann, S.~M., Guo, Y., \& Jing, 
Y.~P.\ 2006, \mnras, 369, 1293 
\end{thebibliography}
\end{document}